\documentclass[fleqn,12pt,twoside]{article}
\usepackage{espcrc1}

\usepackage{graphicx}
\usepackage[figuresright]{rotating}

\newcommand{\AmS}{{\protect\the\textfont2
  A\kern-.1667em\lower.5ex\hbox{M}\kern-.125emS}}

\title{The convective Urca process}

\author{P. Lesaffre\address[IoA]{Institute of Astronomy, Madingley road, Cambridge CB3~0HA, United Kingdom},
Ph. Podsiadlowski\address{University of Oxford, Department of Astrophysics,
 Oxford OX1~3RH, United Kingdom}, and
C.A. Tout\addressmark[IoA]
       }

\begin{document}

\maketitle

\begin{abstract}

 One possible fate of an accreting white dwarf is explosion
 in a type Ia supernova. However, the route to the thermonuclear
 runaway has always been uncertain owing to the lack of a convective
 model consistent with the Urca process.

 We derive a formalism for convective motions involving two radial
 flows. This formalism provides a framework for convective models that
 guarantees self-consistency for chemistry and energy budget, allows
 time-dependence and describes the interaction of convective motions
 with the global contraction or expansion of the star.  In the
 one-stream limit, we reproduce several already existing convective
 models and allow them to treat chemistry.  We also suggest as a
 model easy to implement in a stellar evolution code.

 We apply this formalism to convective Urca cores in Chandrasekhar
 mass white dwarfs.  We stress that in degenerate matter, nuclear
 reactions that change the number of electrons strongly influence the
 convective velocities.  We point out the sensitivity of the energy
 budget on the mixing. We illustrate our model by computing {\it
 stationary} convective cores with Urca nuclei. We show that even a
 very small mass fraction of Urca nuclei ($10^{-8}$) strongly
 influences the convective velocities.

 Finally, we present preliminary computations of the late evolution of
 a close to Chandrasekhar mass C+O white dwarf including the
 convective Urca process.
\end{abstract}

\section{Prelude to thermonuclear explosions}

  When a CO white dwarf accretes matter, its centre gets denser and
  hotter.  At some point, the C-C burning ignites mildly in
  the centre. When the radiative luminosity can no longer get rid of
  the heat produced, a convective core forms and grows as the burning
  releases more and more energy. During the burning process, Urca
  pairs such as $^{23}{\rm Na-}^{23}$Ne are produced. These release
  neutrinos through emission and capture of electrons preferentially
  around Urca shells. The net amount of energy released and the change
  in the electron fraction at the time of the explosion have always
  been uncertain due to the lack of a convective model self-consistent
  with the energy and chemistry budgets
  \cite{P72,B73,CA75,I78a,I78b,I82,BW90,M96,S99,B01}. Former theoretical attempts
  \cite{E83,G93,C99} to describe the interplay between convection
  and chemistry have all left open the question of energy
  conservation.\\

   A better understanding of the phase immediately preceding
  the thermal runaway provides an essential link between the core
  evolution in the progenitor and the supernova explosion. Knowledge
  of the physical and chemical profile at the time of the explosion may
  also provide new insight in the physics of the explosion itself.
  Indeed, the neutron excess at the time of ignition is a crucial
  parameter of the explosion, and it is strongly affected by the
  convective Urca process.

\section{A model for the convective Urca process}

  Except for the work of Iben \cite{I78a,I78b,I82}, the convective
  Urca process has been so far neglected in presupernova
  computations. This is because we lack a proper energy equation in a
  convective region.  Indeed, far from the Urca shell, out of
  equilibrium nuclear heating overwhelms neutrino losses
  \cite{B73}. Hence the net heating crucially depends on the actual
  chemical mixing.  The amount of convective work necessary to carry
  the electrons is uncertain \cite{BW90,M96,S99}. Last but not least,
  in degenerate matter chemistry controls the buoyancy and thence the
  convective velocities.  Fortunately, the problem is well known and
  solved in the {\it radiative} state. Our basic idea is to mimic
  convection by considering the interaction between two {\it
  radiative} columns of fluid side by side \cite{L04}.

 The resulting convective model {\bf guarantees energy
 conservation}. Besides, it is time-dependent, non-local, allows for
 different rising and descending convective velocities, describes
 horizontal chemical inhomogeneities and treats the interaction of
 convection with the mean flow. We have checked carefully different limits
 of our model against corresponding existing convective models
 \cite{E83,G93,U67} and found good agreement.

  Although the full model would be rather difficult to implement in a
  convective core, the one-stream, local, time independent limit
  yields a straightforward extension of the mixing length theory fully
  consistent with chemistry and energy budget. The main difference
  with the mixing length is the dependence of the convective
  velocity on the chemical gradients.

\section{Application to degenerate CO convective cores}

   We use our convective model to compute convective cores with a homogeneous
total abundance for the Urca pair A=23. Indeed this pair operates at mass
density 1.7~10$^9$~g~cm$^{-3}$ and is hence one of the main actors
at the onset of C burning which occurs around density 2~10$^9$~g~cm$^{-3}$
and temperature 3~10$^{8}$~K.

\subsection{Stationary models}
\begin{figure}[h]
\centerline{\includegraphics[angle=0,width=25pc]{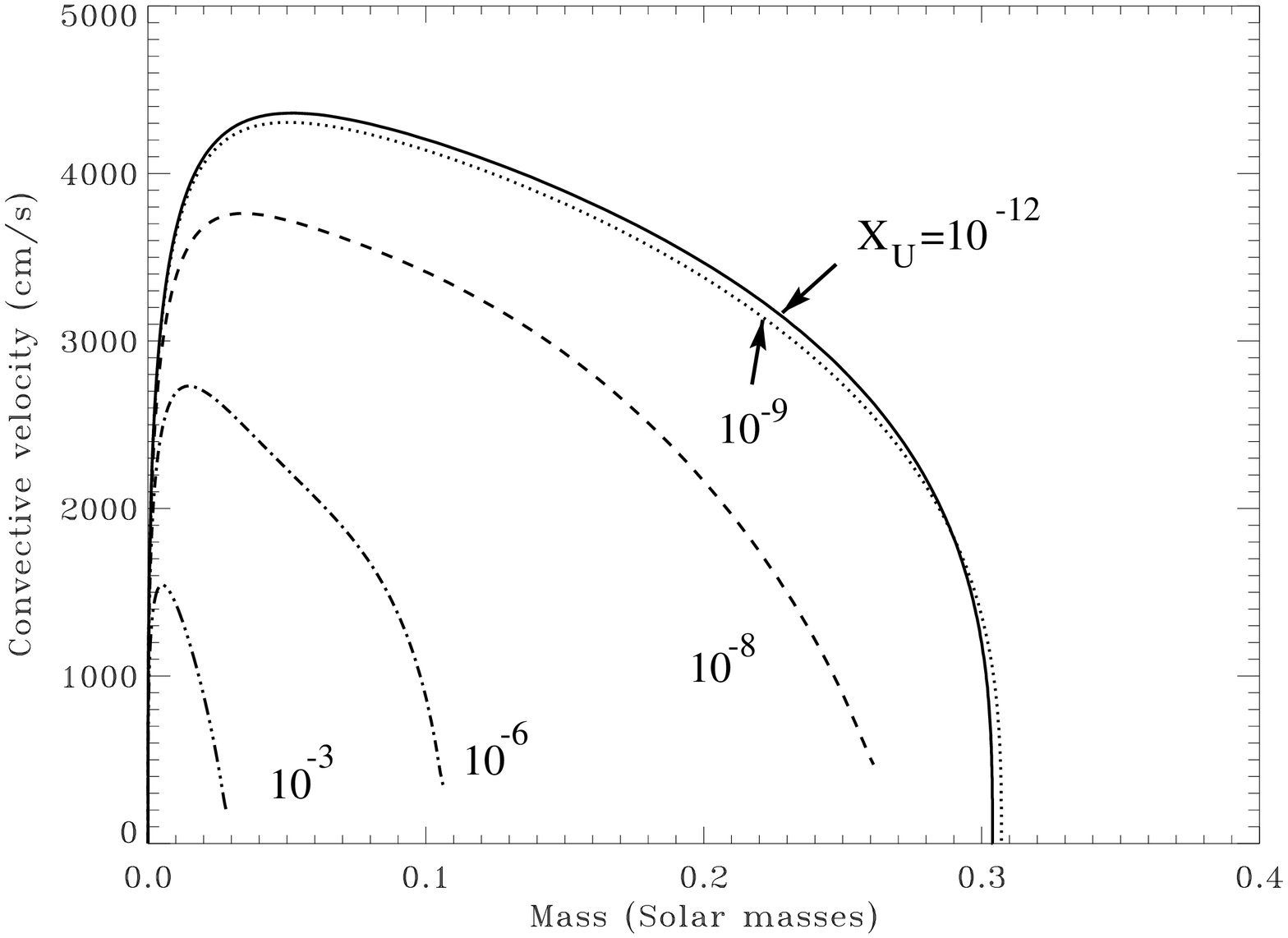}}
 \caption{Variation of the stationary convective profile for different
 mass fractions $X_{\rm U}$ of the Urca pair A=23.}
\label{fig1}
\end{figure}

  We first used the full two-stream model to compute {\it stationary}
  convective cores. We varied the mass fraction of Urca matter $X_{\rm
  U}$ from 10$^{-12}$ to 10$^{-3}$. Strong modifications of the
  convective profile occur for mass fractions higher than 10$^{-8}$ as
  Fig. \ref{fig1} shows.  Analytical arguments on the existence of a
  stationary convective core confirm this low threshold. 

 The convective velocities are strongly reduced by the presence of
  Urca matter. Indeed, the velocity of a convective blob is
  proportional to its difference in density with the surrounding
  medium. Besides, the density of degenerate matter depends mainly on
  the electron number density. This means that {\bf convection in
  degenerate matter crucially depends on electron emissions and
  captures}.
  These simulations also show that the one-stream, local limit is a
  good approximation to the two-stream model in this context.

\subsection{Stellar evolution models}

 We have started to implement a one-stream, stationary limit of the
 two-stream model in P.~Eggleton's stellar evolution code
 \cite{E71}. This has already required significant changes to the code to
 handle chemical gradients with sufficient accuracy. 

  First, it appears that the backreaction of the chemical gradients
on the convective velocity, and hence the mixing, causes serious convergence
problems. We momentarily overcome these problems by artificially lowering 
the mixing at the outer edge of the convective core. Simulations with
$X_{\rm U}=10^{-7}$ confirm the reduction of the convective speed which
now takes place only upward of the Urca shell.

  Second, for higher Urca mass fractions, our code breaks down when the
  outer edge of the convective core reaches the Urca shell. At this point,
the convective state changes on short time scales due to the gradient of
the Urca composition.

\section{Conclusion and future prospects}

We have developed a new convective model suitable for studying the convective
Urca process. A first application to stationary convective cores
suggests that electron emissions and captures control convection in
degenerate matter.

  We hence expect to uncover new results on the initial conditions for
  the explosion of type Ia (and possibly core-collapse supernovae). 
 In particular, we shall be able to predict the degree of
  neutronisation at the time of the thermal run-away, a
  crucial parameter for explosive nucleosynthesis

  However, we still need to implement a fully time-dependent version
  of our convective model in a stellar evolution code including an
  extended chemical network.

  Finally, our model may well prove to be relevant for other fields of
  stellar evolution such as the He flash, the fate of O, Ne and Mg cores
  and AGB stars.

\end{document}